\documentclass{article}                             
\usepackage{frascatiphys,here,graphicx,subfigure}
\usepackage{relsize}         
\RequirePackage{xspace}
\def\babar{\mbox{\slshape B\kern-0.1em{\smaller A}\kern-0.1em         
B\kern-0.1em{\smaller A\kern-0.2em R}}}         

\mathchardef\Upsilon="7107
\def\Y#1S{\ensuremath{\Upsilon{(#1S)}}\xspace}

\def\pep2{PEP-II}

\mathchardef\Lambda="7103
\mathchardef\Sigma="7106
\def\Sigbar{\ensuremath{\kern 0.2em\overline{\kern -0.2em \Sigma}}{}}
\def\Lbar {\ensuremath{\kern 0.2em\overline{\kern -0.2em\Lambda\kern 0.05em}\kern-0.05em}{}}
\def\proton      {\ensuremath{p}\xspace}
\def\antiproton  {\ensuremath{\overline p}\xspace}
\def\B       {\ensuremath{B}\xspace}
\def\Bbar    {\kern 0.18em\overline{\kern -0.18em B}{}\xspace}

\begin{document}
\title{\bf Studies of Exclusive $e^+e^-\to~ hadrons$
Reactions with Baryons and
Strange Particles Using Initial State Radiation at Babar}
\author{S.I.Serednyakov, for Babar Collaboration\\ 
\em{\it Budker Institute of Nuclear Physics,}\\
\em{\it Novosibirsk State University,}\\
\em{\it 630090, Novosibirsk, Russia} }
\maketitle
\baselineskip=11.6pt 
\begin{abstract}  New  \babar\  results on exclusive  $e^+e^-\to~ hadrons$
reactions at low c.m. energy with a baryon-antibaryon 
or $K\bar{K}$  pair in the final 
state are presented. Cross sections are measured using the 
initial state radiation technique  from the threshold up to 4--5 GeV. 
From the measured
$e^+e^-\to\proton\antiproton,~ \Lambda\Lbar,~\Sigma^0\Sigbar^0,~
\Lambda\Sigbar^0 (\Sigma^0\Lbar)$ cross sections  we derive 
effective baryon form factors and compare them with predictions. 
We measure the 
$K\bar{K}\pi (\eta),~K^+K^-\pi\pi, ~K^+K^-3\pi,~K^+K^-4\pi, ~K^+K^-K^+K^-$ 
final states and also study  their internal structure. 
A new $Y(2175)$ state is observed  in $e^+e^-\to~K^+K^-f_0(980)$, with 
$f_0\to\pi\pi$.  The total  measured $e^+e^-$        
annihilation cross section into final states including 
strange baryons or  strange mesons   
is estimated to be 10\% of the full
hadronic cross section.
\end{abstract}

\baselineskip=14pt
{\bf Introduction.} The  $e^+e^-\to~ hadrons$ reactions with a
pair of baryons
or  strange mesons in the final state are the subject of the 
experimental study  for many years for  several reasons. 
First, they give considerable contribution into the
total $e^+e^-$ hadronic cross section. Second, 
from the two body cross section  one can
derive electromagnetic timelike form factors. And last but not least, 
new states can reveal themselves in the study of these reactions. 
   
   In this work a sample of \babar\ \cite{babar}      
data corresponding to 230 $fb^{-1}$  is analyzed.
We search for initial state radiation (ISR) processes            
$e^+e^- \to f + \gamma$,  where  $\gamma$ is a  high energy photon with             
$E_{\gamma}>3$ GeV and $f$ is a hadronic system with the mass $m$.
(A description of ISR approach is given in  \cite{ppbar}).
Through ISR the wide mass range            
(from $2m_{\pi}$ to $\sim 10~GeV/c^2$) is studied in a single experiment     
with full efficiency and full  angular acceptance beginning from the very
threshold. For example, in the reaction $e^+e^-\to\proton\antiproton\gamma$,
the protons produced at the threshold already have 
the laboratory momenta  $\geq 1~ GeV/c$.

The ISR approach, applied to B-factories data, is quite            
competitive with direct $e^+e^-$ experiments,            
because the effective  ISR luminosity is            
comparable with already stored   $e^+e^-$ luminosity.
The following final states are studied in this work:            
$\proton\antiproton,~ \Lambda\Lbar,~\Sigma^0\Sigbar^0,~            
\Lambda\Sigbar^0 (\Sigma^0\Lbar)$ with baryons and            
$K\bar{K}\pi (\eta),~K^+K^-\pi\pi,$  $K^+K^-3\pi,$            
$K^+K^-4\pi, ~K^+K^-K^+K^-$ and other with kaons.            
Other ISR results are covered in the talk \cite{Weng}.

{\bf Baryon pair production results.}
 The cross section for the $e^+e^- \to \B\Bbar$ process, where \B\ is
a spin-1/2 baryon has the form:
\begin{equation} 
\sigma_{\B\Bbar}(m) = \frac{4\pi\alpha^{2}\beta}{3m^2} 
\left [|G_M(m)|^{2} + \frac{1}{2\tau}|G_E(m)|^{2}\right], 
\label{eq4} 
\end{equation} 
where $\beta =\sqrt{1-4m_B^2/m^2}$,  $\tau=m^2/4m_B^2$, 
$G_E$ and  $G_M$ are the electric and magnetic form factors.
From the total cross section (\ref{eq4}) the values 
of the $G_E$ and  $G_M$ 
can not be extracted separately. 
Therefore, the effective form factor $|F(m)|$ is introduced  as 
$|F(m)|^2=(2\tau|G_M(m)|^{2} + |G_E(m)|^{2}/(2\tau+1)$. 
The modulus $|G_E/G_M|$ is determined from the $\cos{\theta_B}$ distribution,
where $\theta_B$ is the polar angle of the baryon with respect 
to the $e^-$ beam in the $e^+e^-$ c.m. frame. The
$\sin^{2}\theta_B$ term in this distribution is 
proportional to  $|G_E|^2$ and the $1+\cos^{2}\theta_B$
term - to $|G_M|^2$.  Fitting 
of the $\cos{\theta_B}$ distribution  then gives the $|G_E/G_M|$ ratio.

\begin{figure}[tbp]
\begin{center}
\begin{minipage}{0.45\textwidth} 
\includegraphics[width=.95\linewidth]{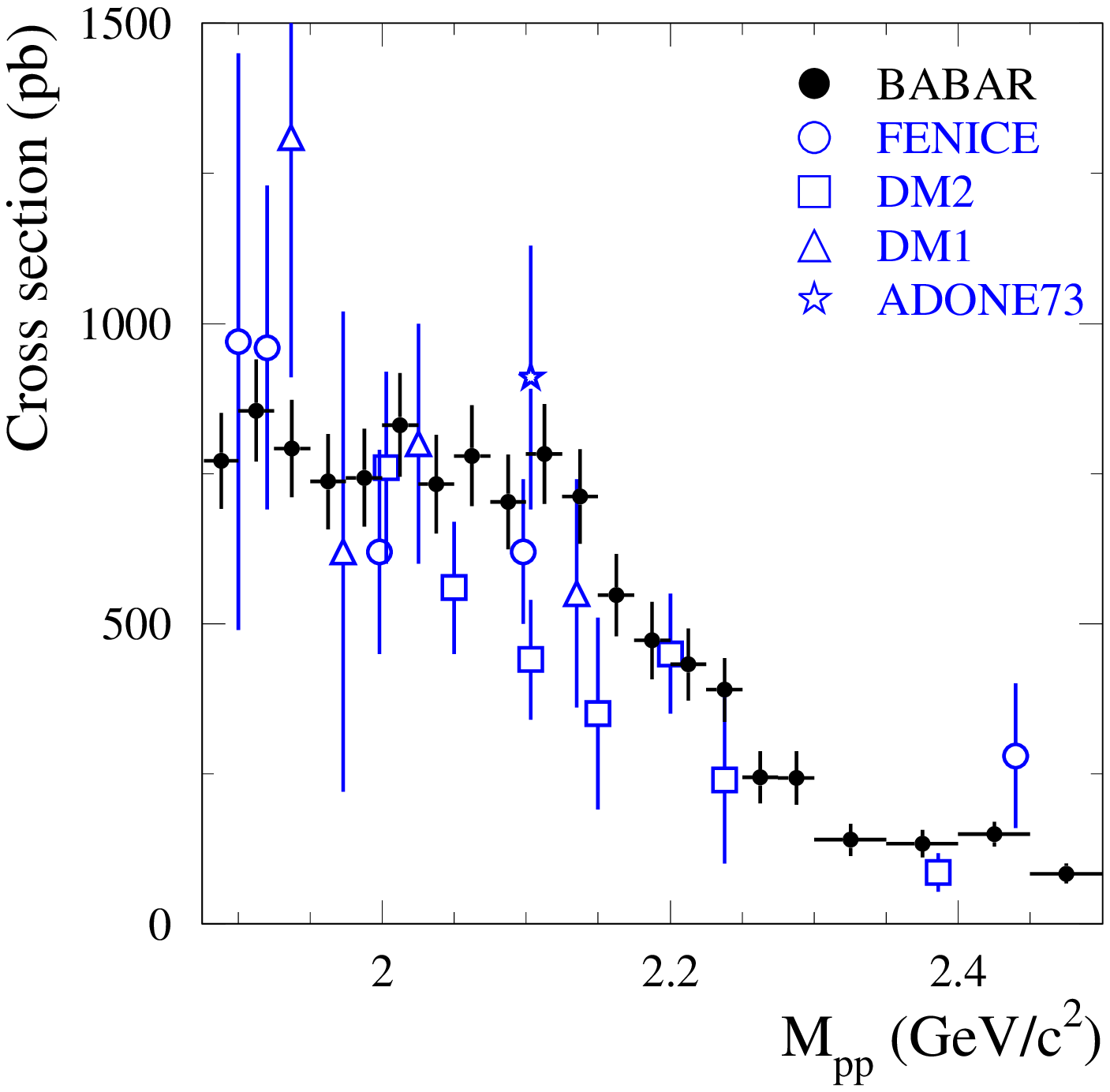} 
\caption{The $e^+e^-\to p\bar{p}$ cross section near threshold.} 
\label{crosspp} 
\end{minipage}
\hfill
\begin{minipage}{0.45\textwidth} 
\includegraphics[width=.95\linewidth]{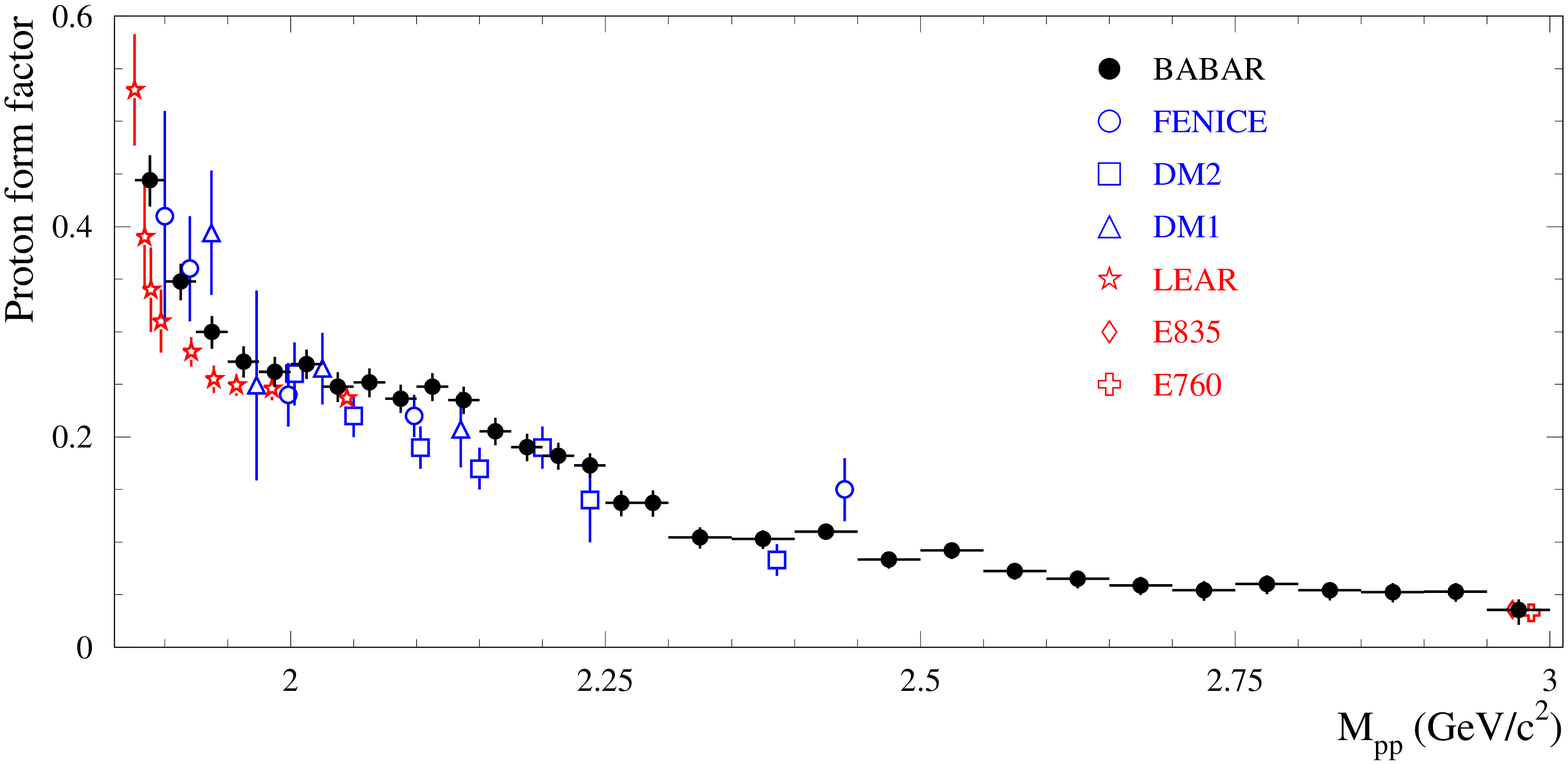} 
\caption{The proton timelike form factor data.} 
\label{ffprot} 
\end{minipage}
\begin{minipage}{0.25\textwidth} 
\includegraphics[width=.95\linewidth]{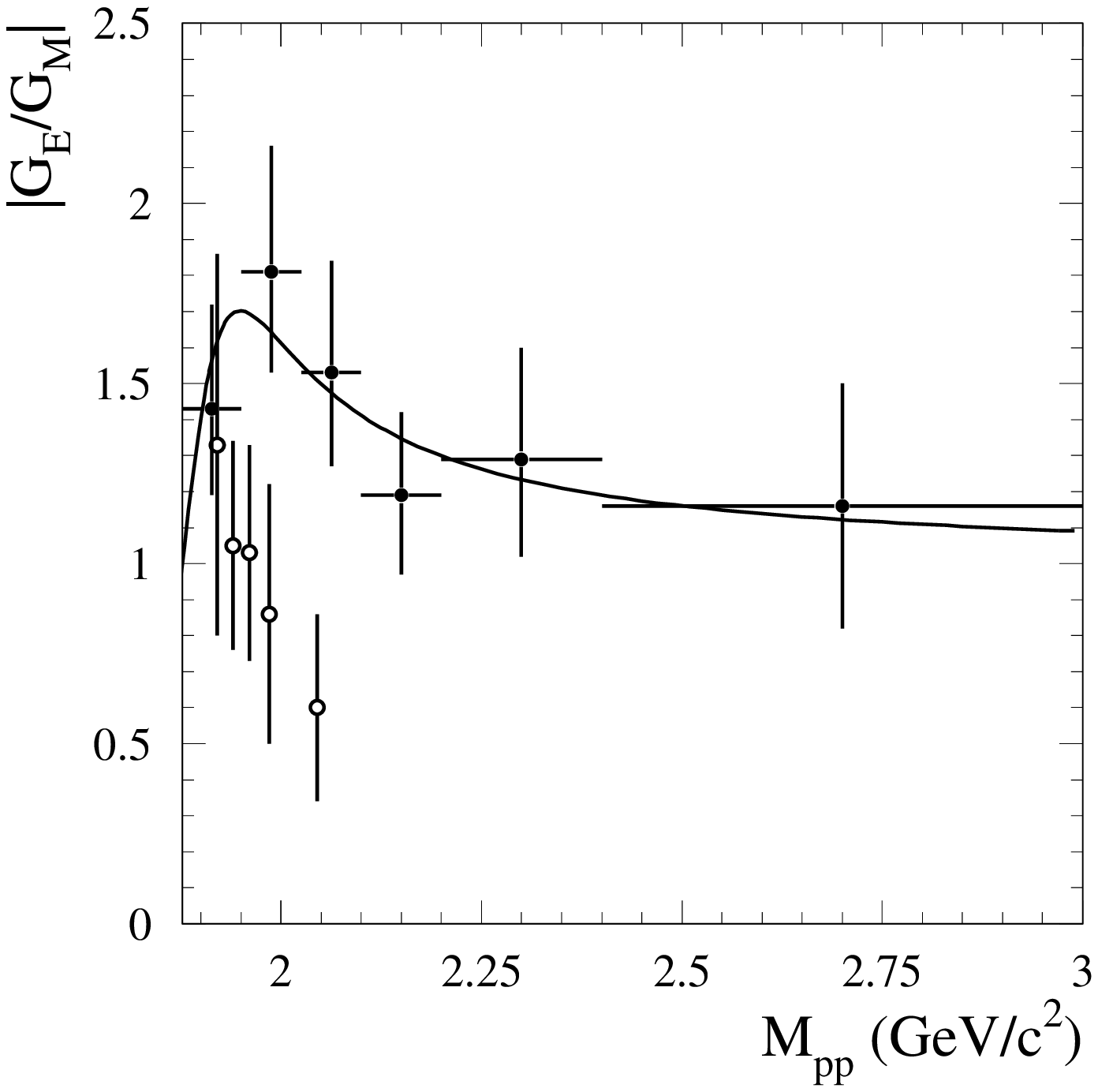} 
\caption{The $G_E/G_M$ ratio for proton.} 
\label{gegmrat} 
\end{minipage}
\hfill
\begin{minipage}{0.27\textwidth} 
\includegraphics[width=.95\linewidth]{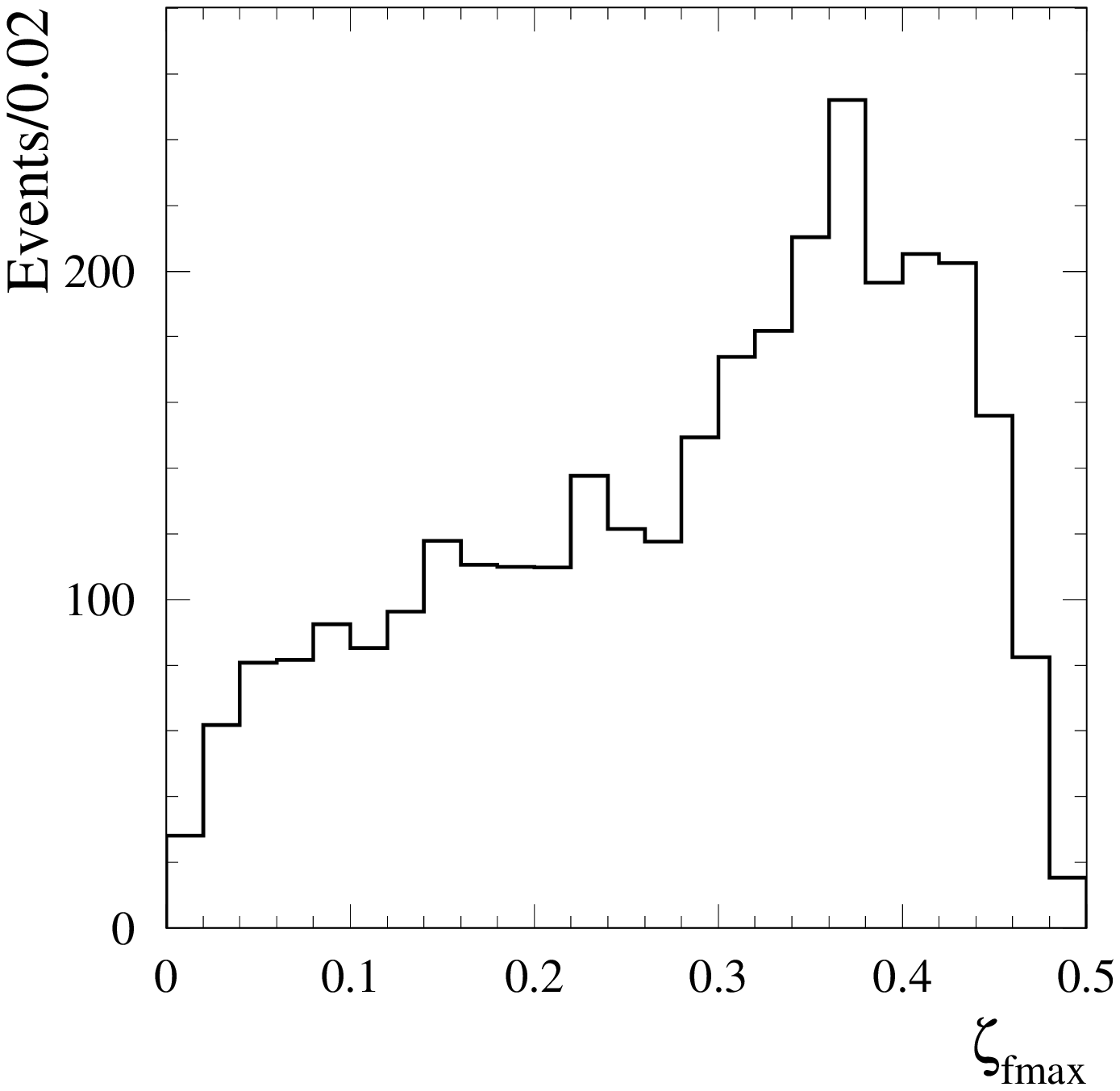} 
\caption{The simulated  $\Lambda$ polarization. } 
\label{zitmax}
\end{minipage}
\hfill
\begin{minipage}{0.27\textwidth} 
\includegraphics[width=.95\linewidth]{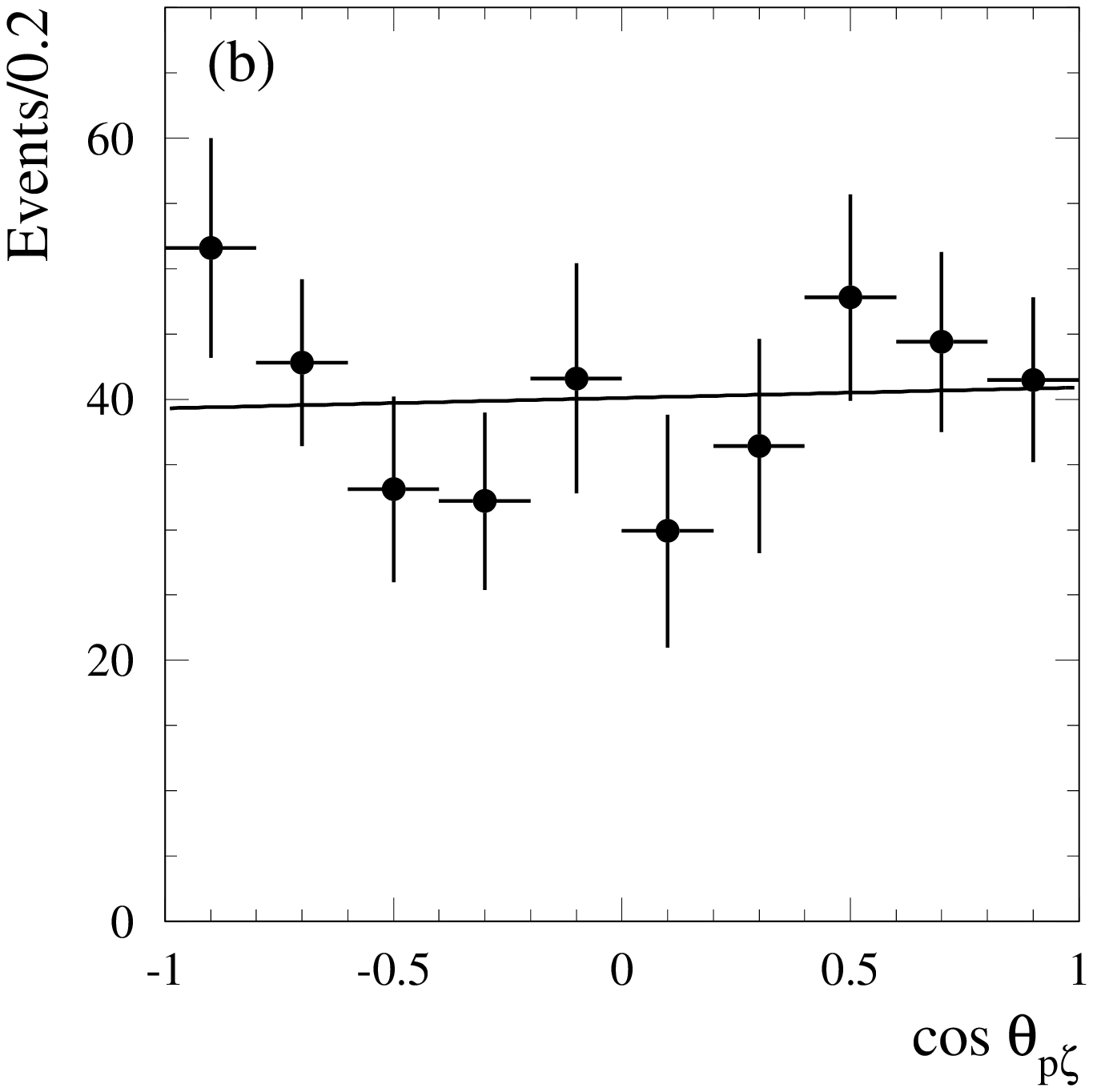} 
\caption{The measured $\cos{\theta_{p\zeta}}$ distribution.} 
\label{coslmbd} 
\end{minipage}
\begin{minipage}{0.42\textwidth} 
\includegraphics[width=.95\linewidth]{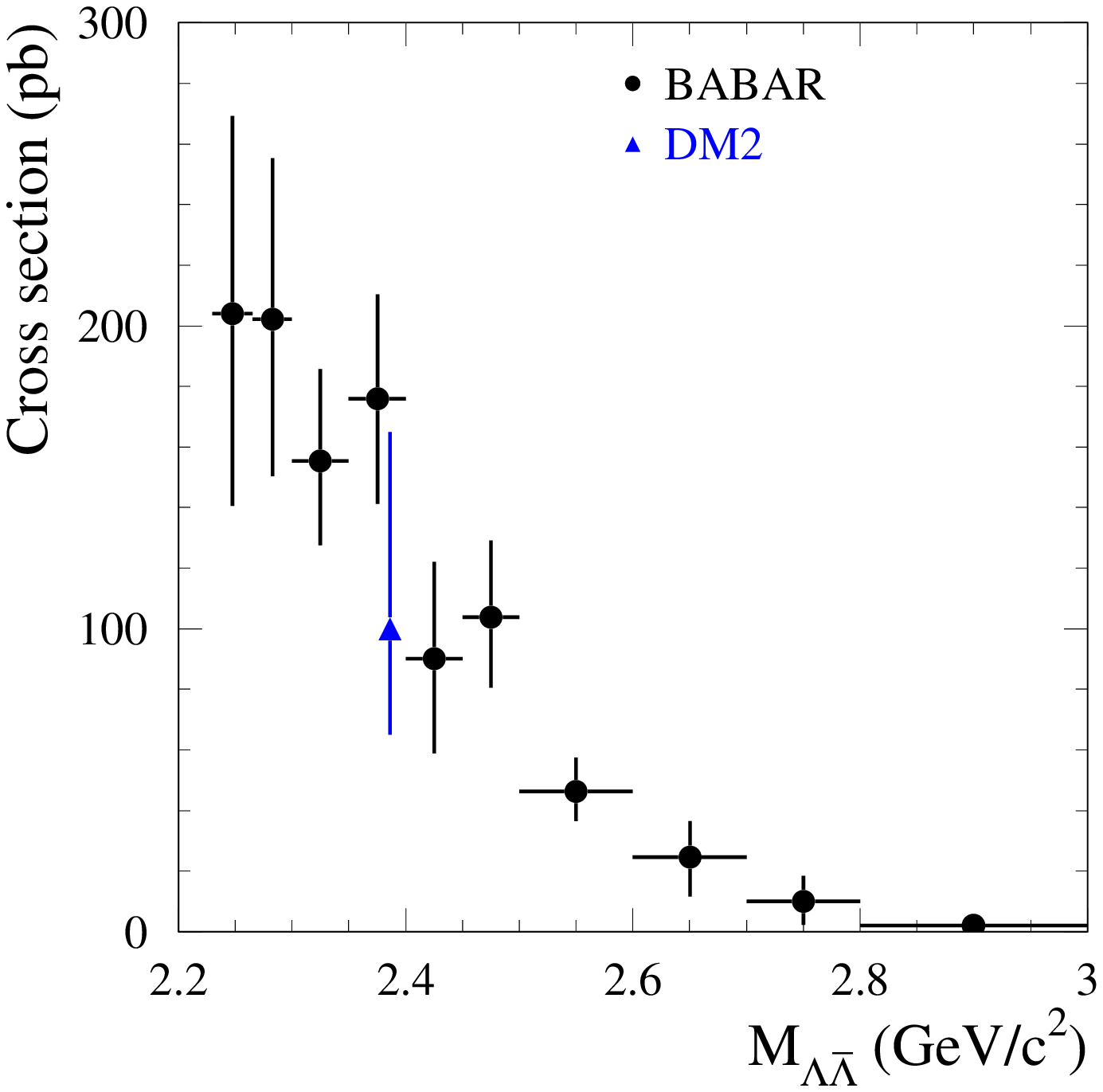} 
\caption{The $e^+e^-\to \Lambda\bar{\Lambda}$ cross section.} 
\label{crosslam} 
\end{minipage}
\hfill
\begin{minipage}{0.42\textwidth} 
\includegraphics[width=.95\linewidth]{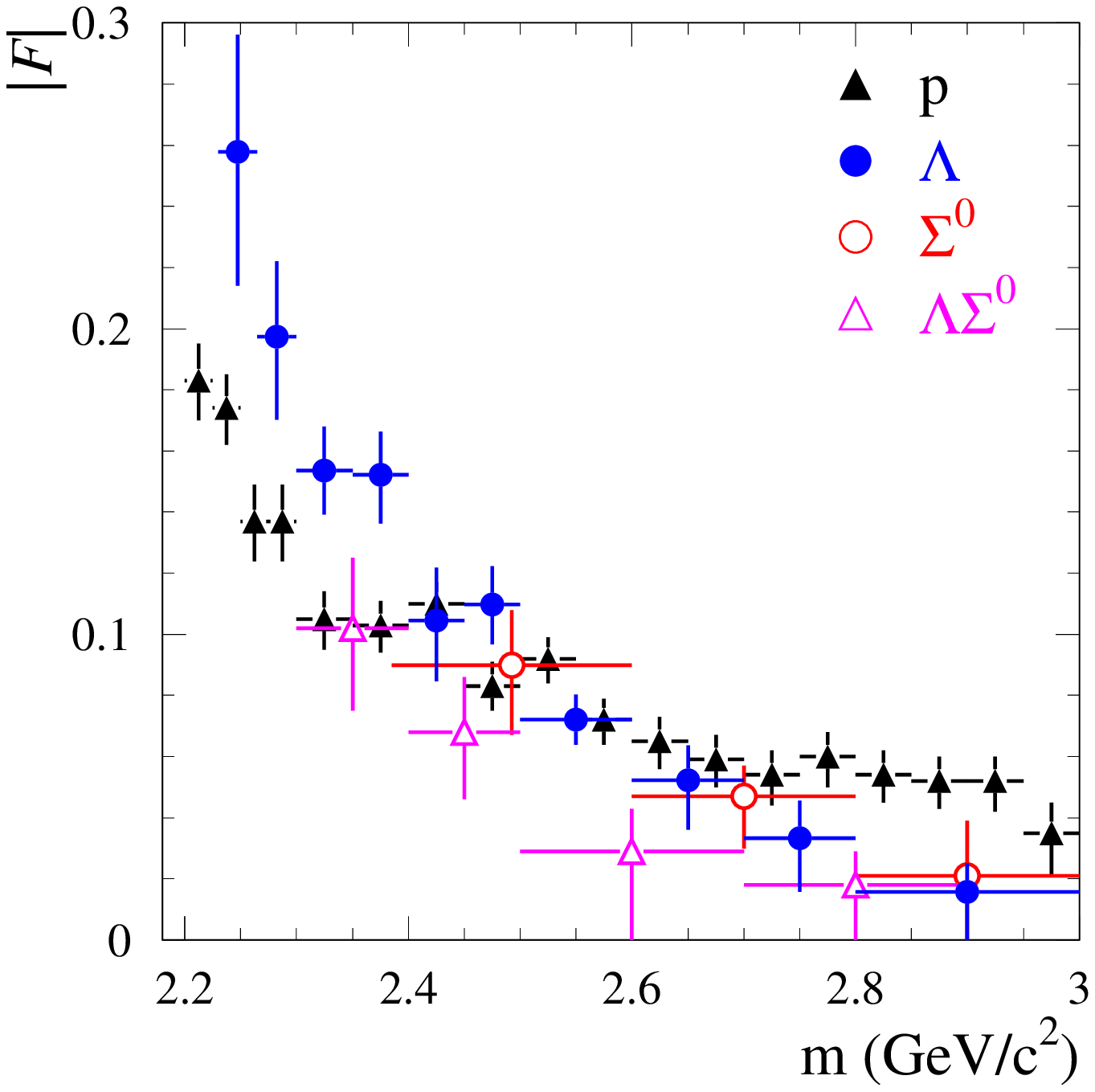} 
\caption{The measured proton, $\Lambda$, $\Sigma^0$ 
and $\Lambda\Sigma^0$  form factors. } 
\label{formall} 
\end{minipage}
\end{center}
\end{figure}

   The $e^+e^-\to p\bar{p}$ results \cite{ppbar} are shown in 
Figs.\ref{crosspp},\ref{ffprot},\ref{gegmrat}. The measured cross section
(Fig.\ref{crosspp}) is flat at the threshold while the  form factor sharply rises
in this region  (Fig.\ref{ffprot}). It  has two step-like structures: 
at 2.15 and 2.9 GeV. The ratio
$|G_E/G_M|$ (Fig.\ref{gegmrat}) is found to be $\geq ~1$ in
contradiction with previous works.

   The measured $e^+e^-\to \Lambda\Lbar$ cross section (Fig.\ref{crosslam})
\cite{LLBar}  agrees 
with the only previous measurement \cite{DM2ll}.  200
found  $\Lambda\Lbar$ events are selected by using  $\Lambda\to p\pi$ decay.
The measured                                
$\Lambda$ effective form factor is shown in Fig.\ref{formall}. The ratio     
$G_E/G_M$ for the $\Lambda$ baryon is consistent with unity.
Use of the  $\Lambda\to p\pi$ decay
allows to measure  the relative phase $\phi$ between 
the complex  $G_E$ and $G_M$ form factors.
The transverse polarization $\zeta$ of  outgoing      
baryons is proportional to  $\sin \phi$.
The measurement of  $\zeta$ is done  from  the 
angular spectrum of protons in the $\Lambda\to p\pi$ decay. 
The simulated distribution over $\zeta_{max}$ is shown in Fig.\ref{zitmax}.
The measured $\cos\theta$ distribution,
where $\theta$ is the angle between the $\Lambda$ polarization and  
the proton momentum from  the $\Lambda\to p\pi$ decay
in the $\Lambda$ rest frame, is shown in
Fig.\ref{coslmbd}. No $\cos\theta$ asymmetry is seen. The following limits
on the $\Lambda$ polarization $-0.22<\zeta<0.28$ and the phase
$-0.76<\sin\phi<0.98$ are obtained. The limit on $\sin\phi$ is too weak 
to make any 
certain conclusion on the phase  $\phi$ between the  $G_E$ and $G_M$
for  the $\Lambda$ hyperon.

    In a similar way the $e^+e^-\to \Sigma^0\Sigbar^0$ and $e^+e^-\to
\Sigma^0\Lbar(\Lambda\Sigbar^0)$ cross sections are measured.
For detection of $ \Sigma^0$, the decay chain $\Sigma^0\to\Lambda\gamma\to
p\pi\gamma$ is used. About 20 candidate events are selected
for each  reaction.   The effective  $\Sigma^0$ 
and $\Sigma^0\Lambda$ form factors are shown in Fig.\ref{formall}. It is seen
that $\Lambda$, $\Sigma^0$ and $\Sigma^0\Lambda$ form factors  
are of the same order.

\begin{figure}[tbp]
\begin{minipage}{0.3\textwidth}  
\includegraphics[width=.95\linewidth]{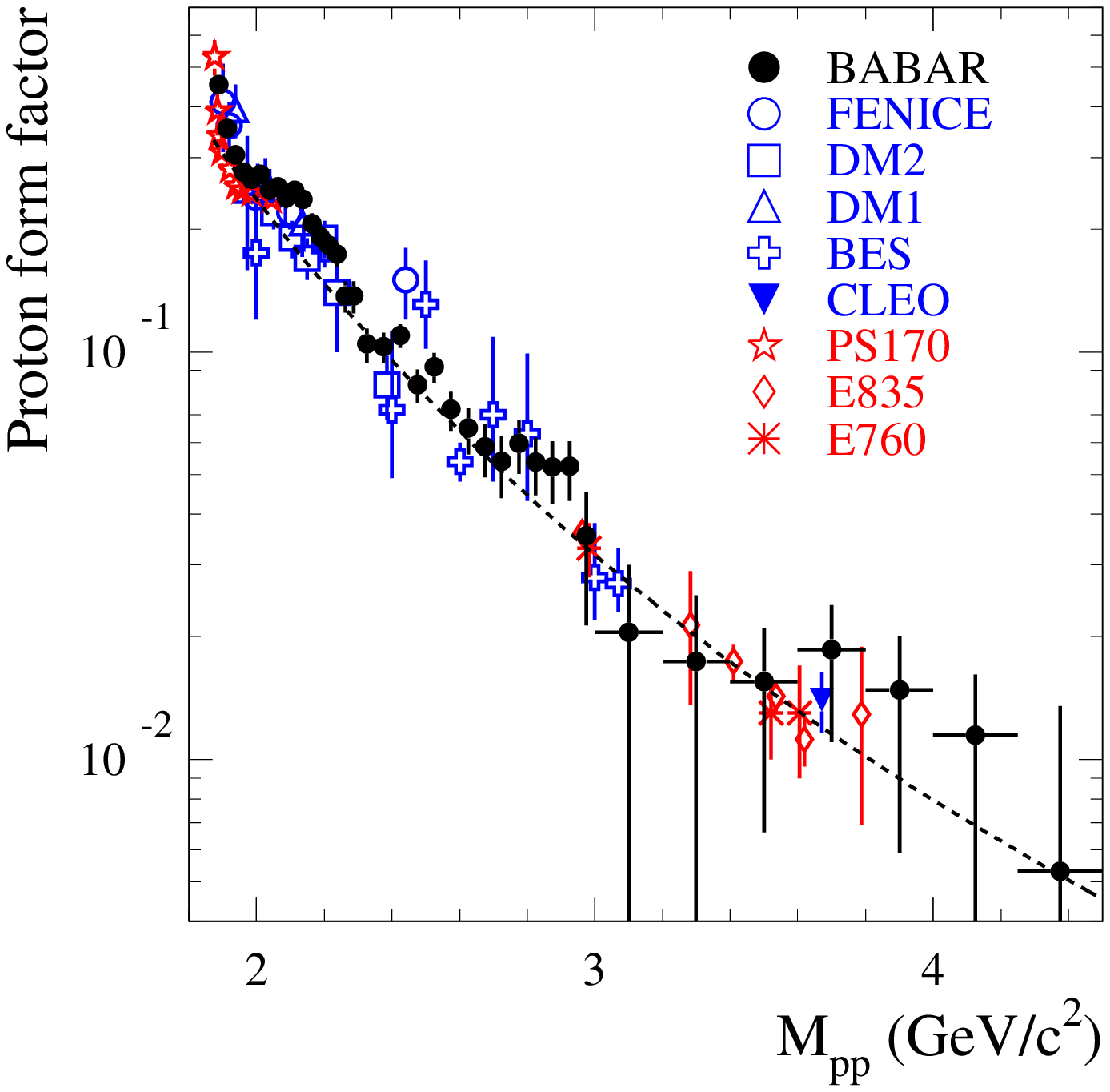}
\caption{The summary of the proton form factor data and QCD fit.} 
\label{qcdprot} 
\end{minipage}
\hfill
\begin{minipage}{0.3\textwidth} 
\includegraphics[width=.95\linewidth]{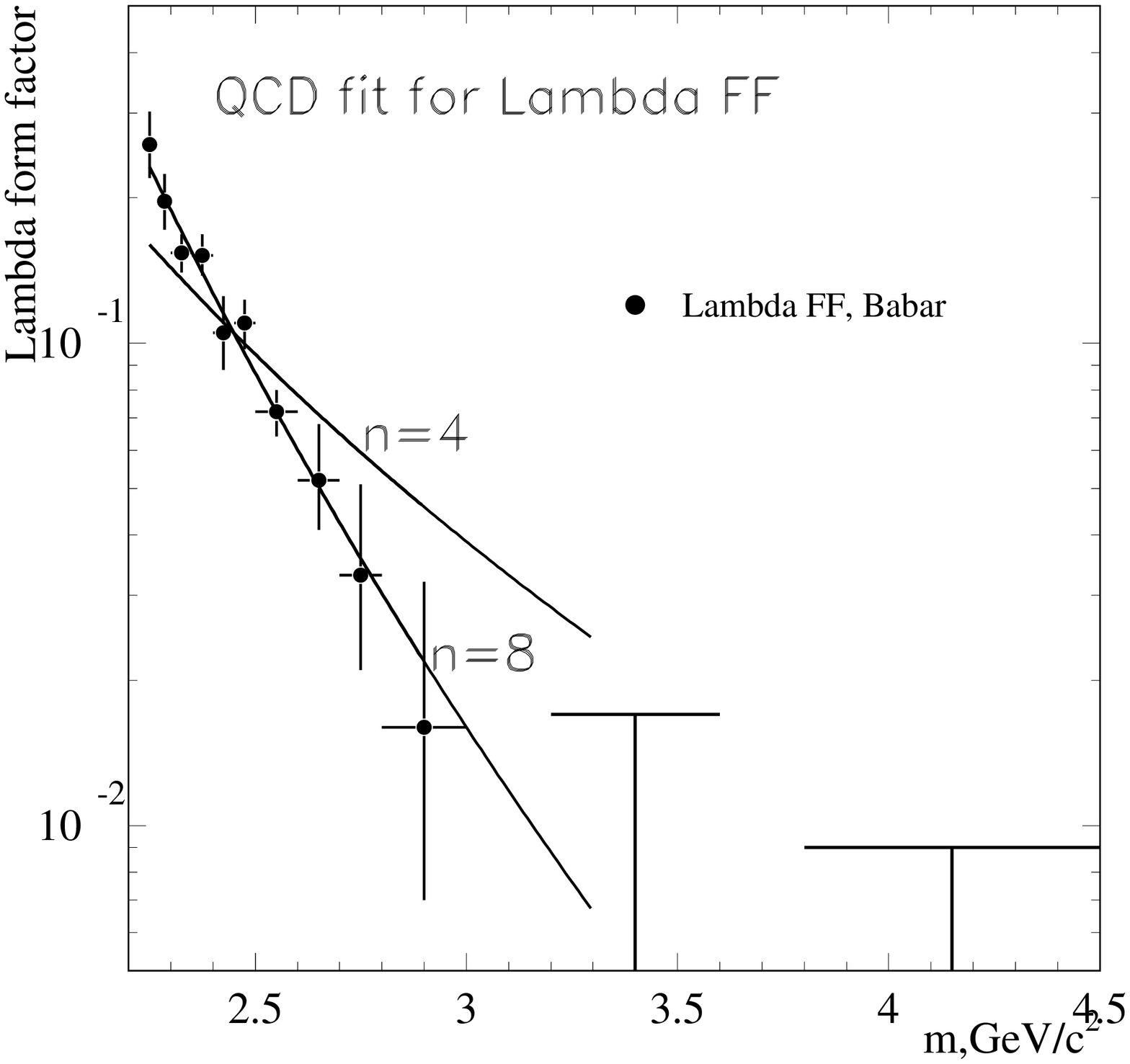}
\caption{ The fitting of  $\Lambda$  form factor data. } 
\label{qcdlam} 
\end{minipage}
\hfill
\begin{minipage}{0.3\textwidth} 
\includegraphics[width=.95\linewidth]{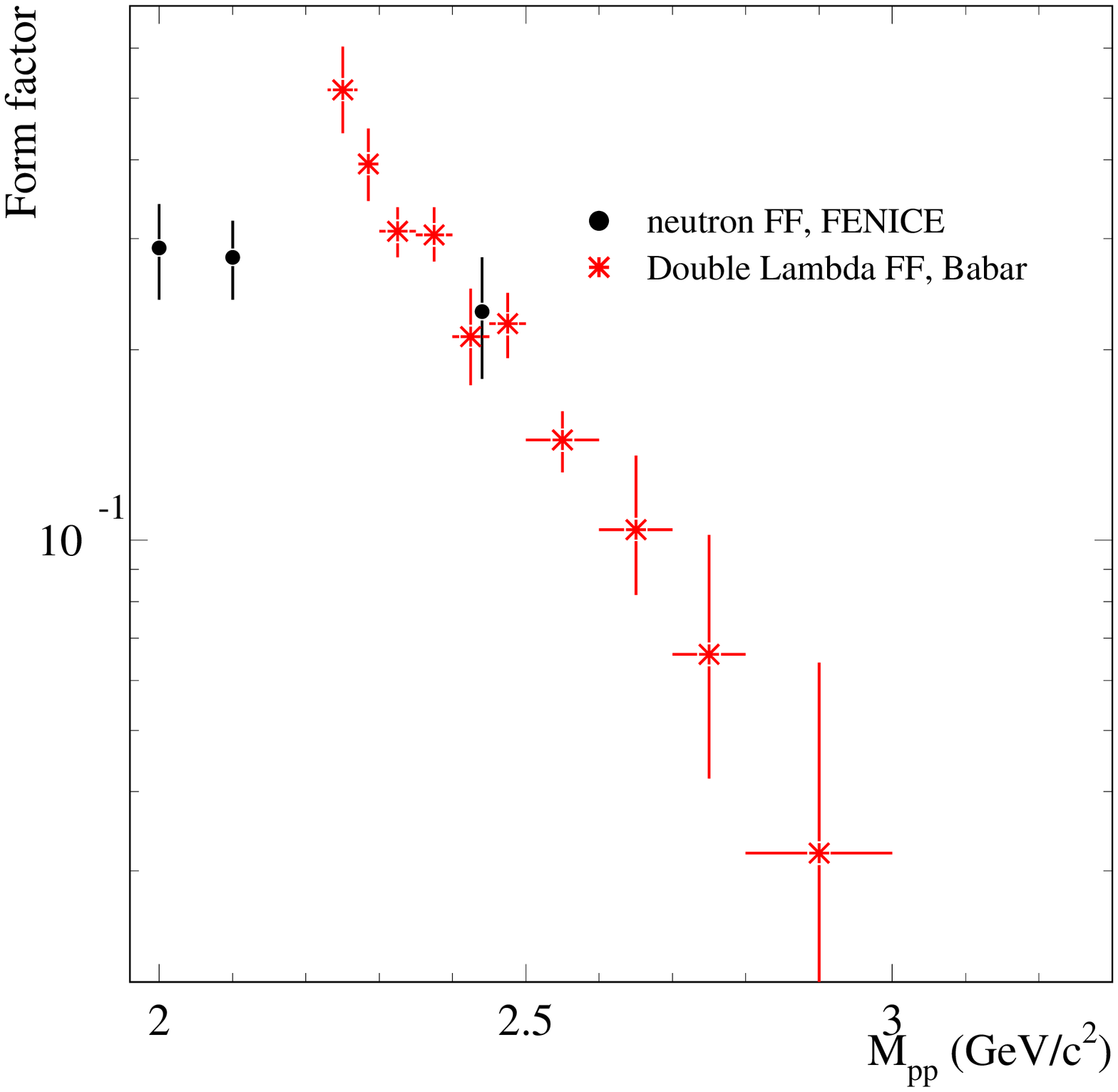} 
\caption{A comparison between doubled $\Lambda$ 
and Fenice neutron \cite{Fenice} form factors.} 
\label{netrlamb} 
\end{minipage} 
\begin{minipage}{0.3\textwidth} 
\includegraphics[width=.95\linewidth]{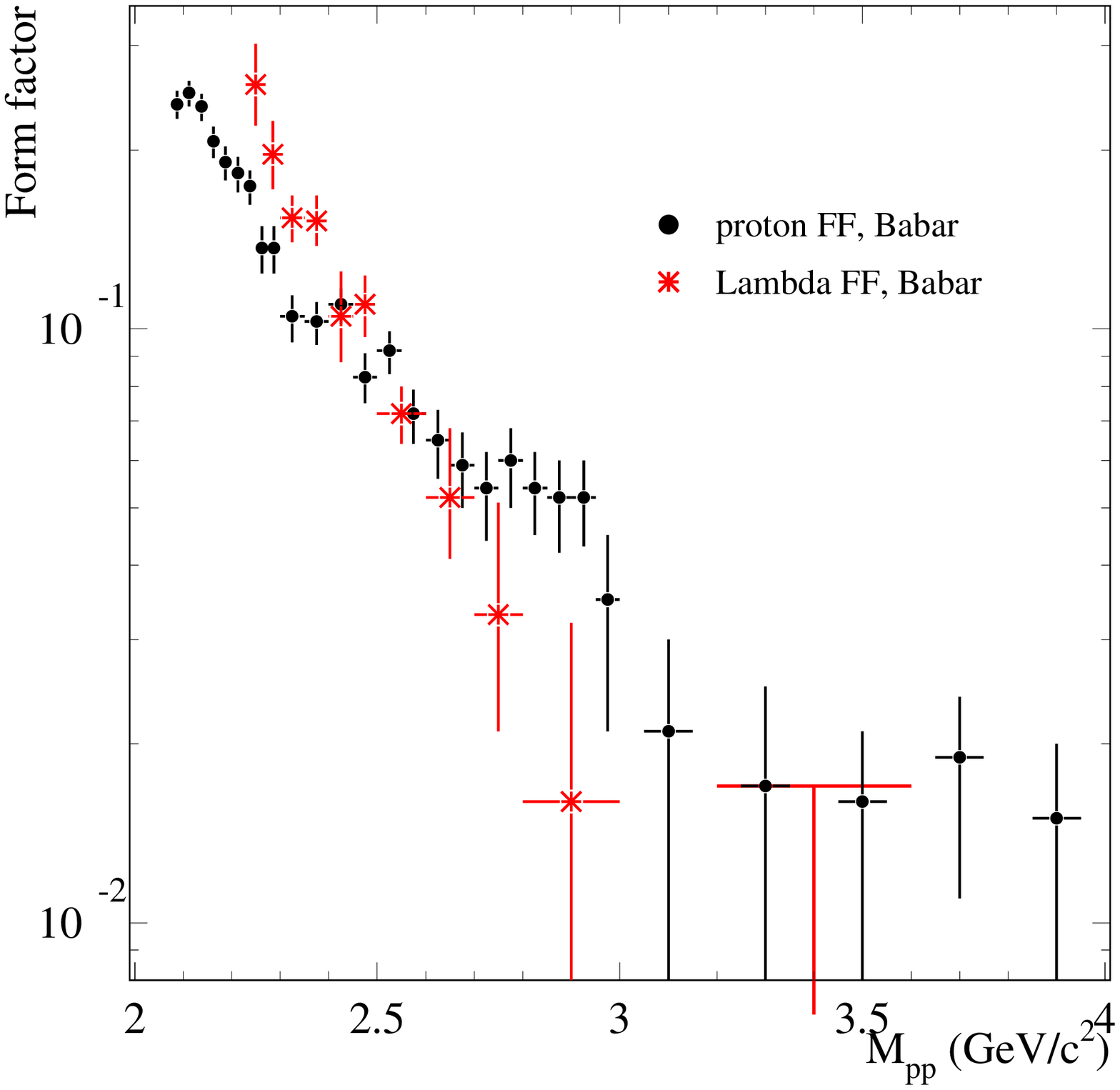} 
\caption{A comparison between proton and $\Lambda$ 
form factors.} 
\label{protlamb} 
\end{minipage} 
\hfill
\begin{minipage}{0.3\textwidth} 
\includegraphics[width=.95\linewidth]{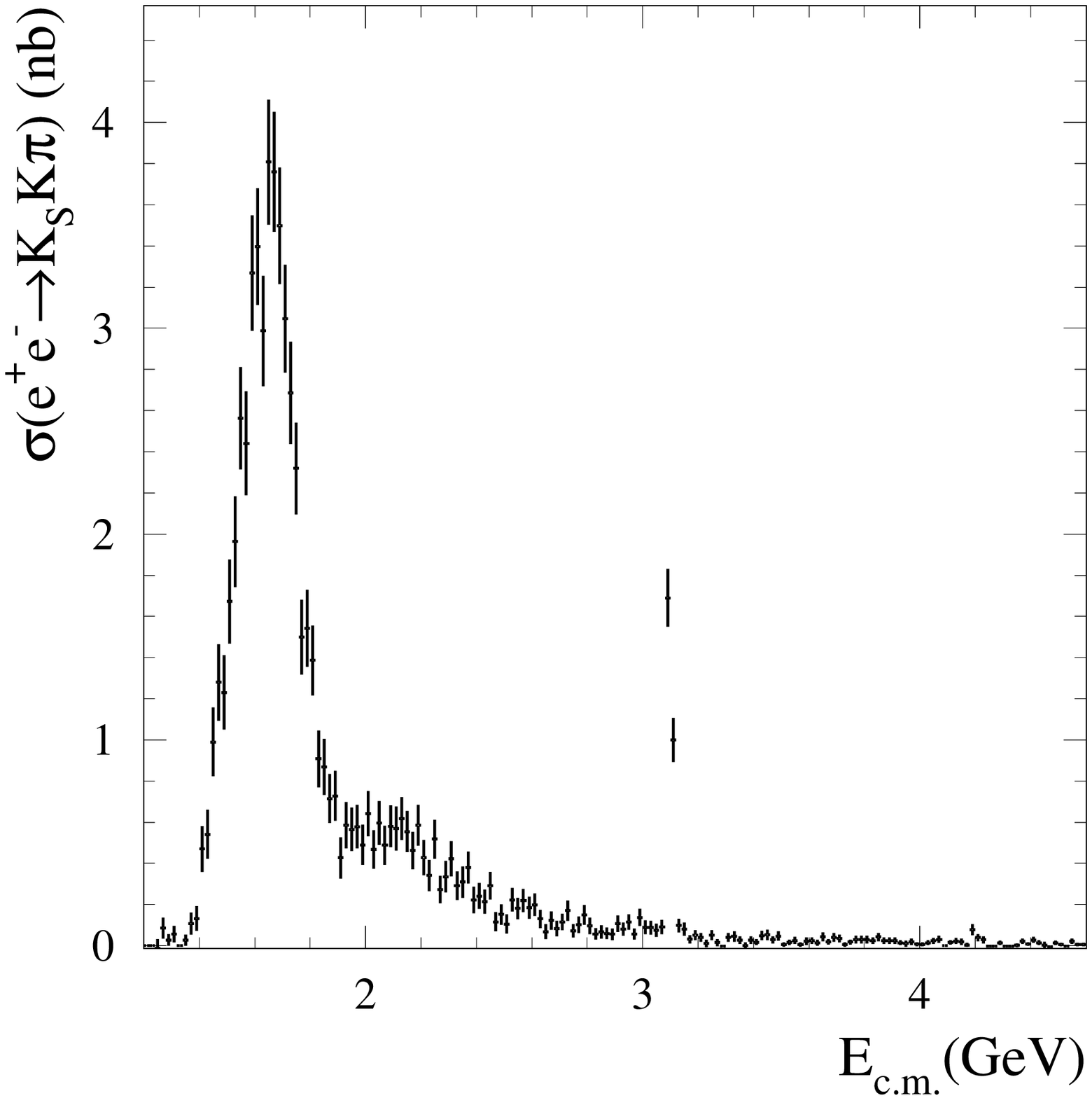} 
\caption{The $e^+e^-\to K_SK^{\pm}\pi^{\mp} $ cross section.} 
\label{kkspi} 
\end{minipage} 
\hfill
\begin{minipage}{0.3\textwidth} 
\includegraphics[width=.95\linewidth]{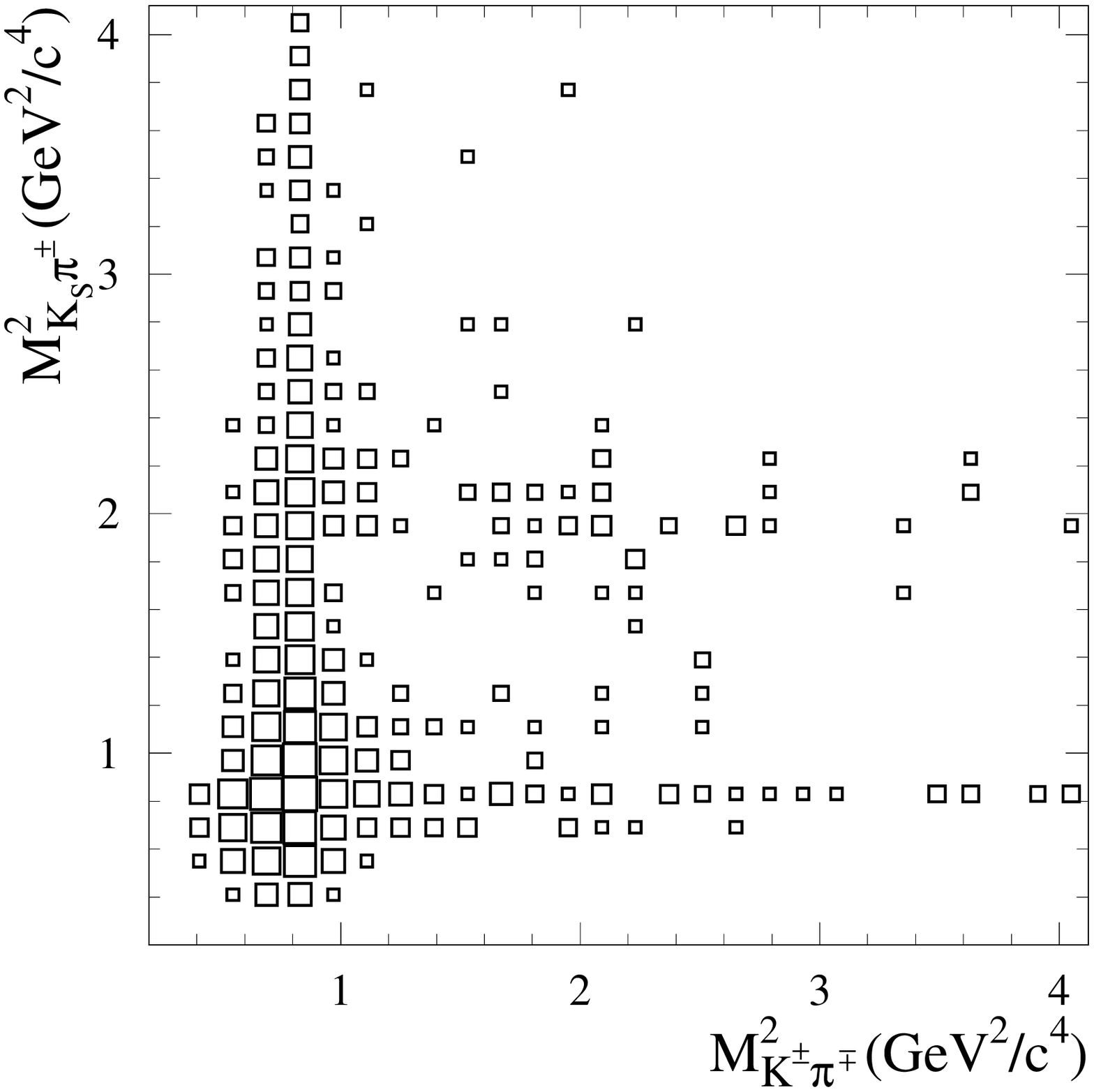} 
\caption{The Dalitz plot for the 
$K_SK^{\pm}\pi^{\mp} $ state. } 
\label{dalkks} 
\end{minipage} 
\begin{minipage}{0.3\textwidth} 
\includegraphics[width=.95\linewidth]{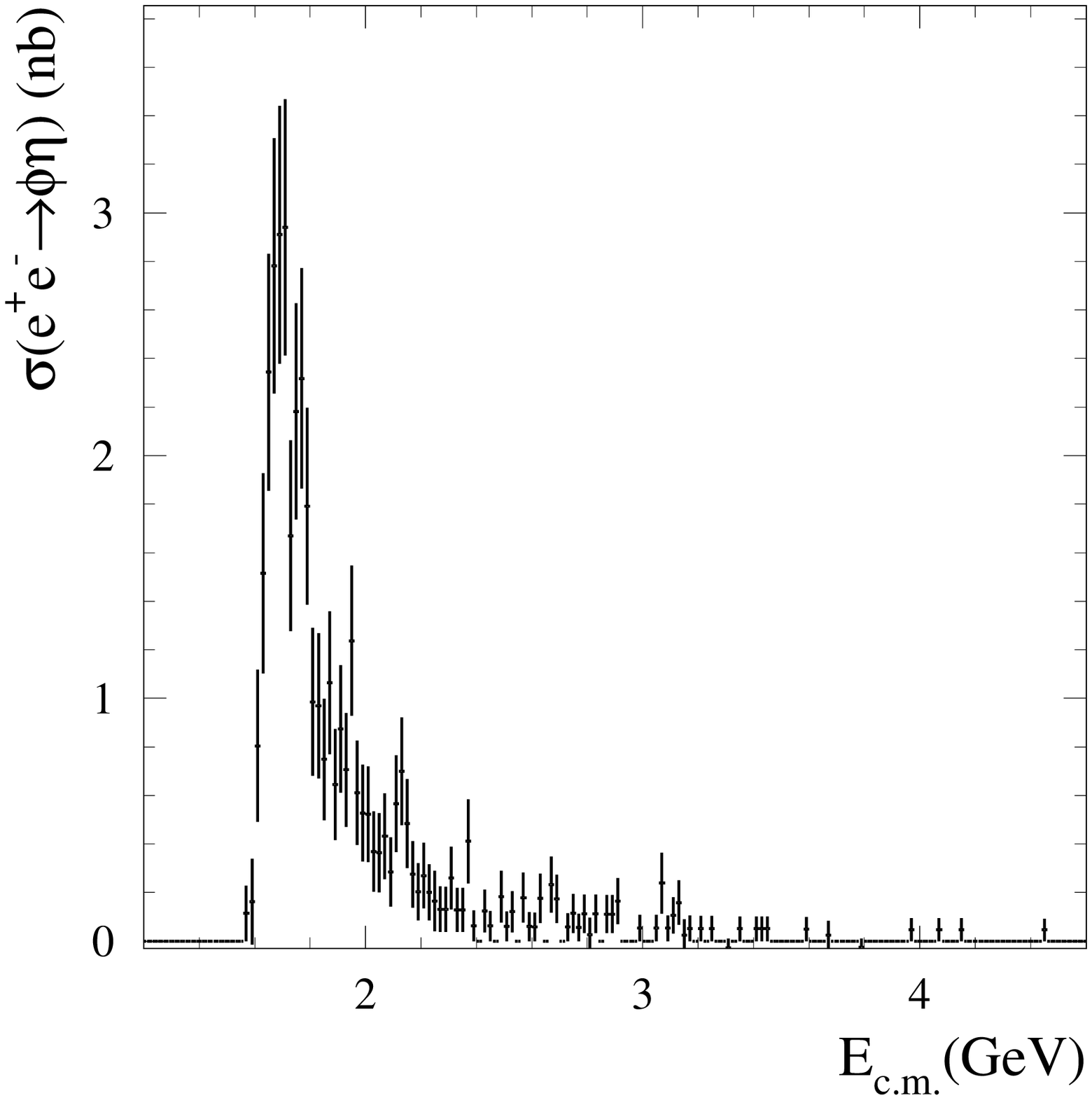} 
\caption{The $e^+e^-\to\phi\eta$  cross section.} 
\label{sigphieta} 
\end{minipage}
\hfill
\begin{minipage}{0.3\textwidth} 
\includegraphics[width=.95\linewidth]{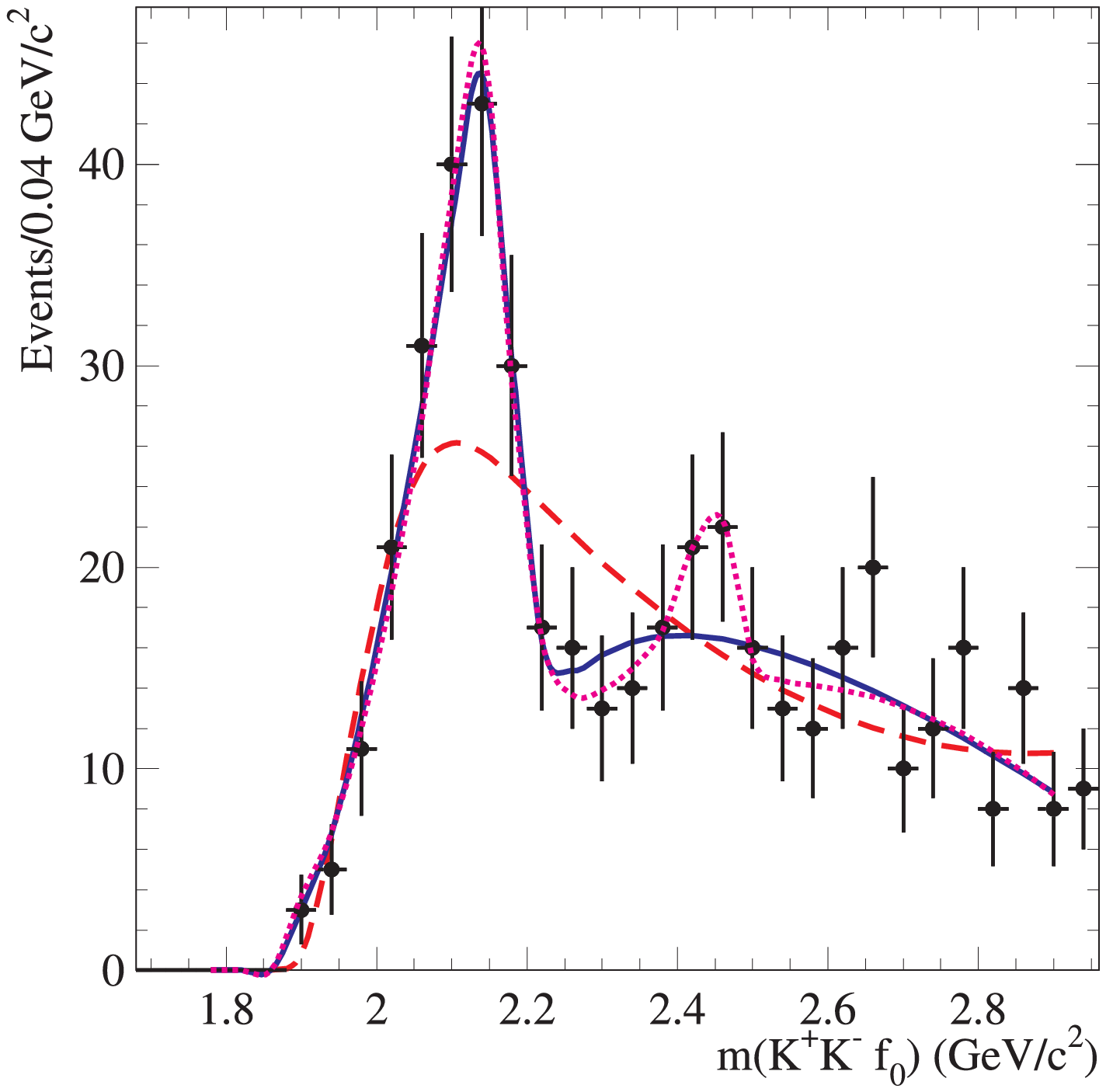} 
\caption{The manifestation of the $X(2175)$ state.}
\label{x2175} 
\end{minipage} 
\hfill 
\begin{minipage}{0.3\textwidth} 
\includegraphics[width=.95\linewidth]{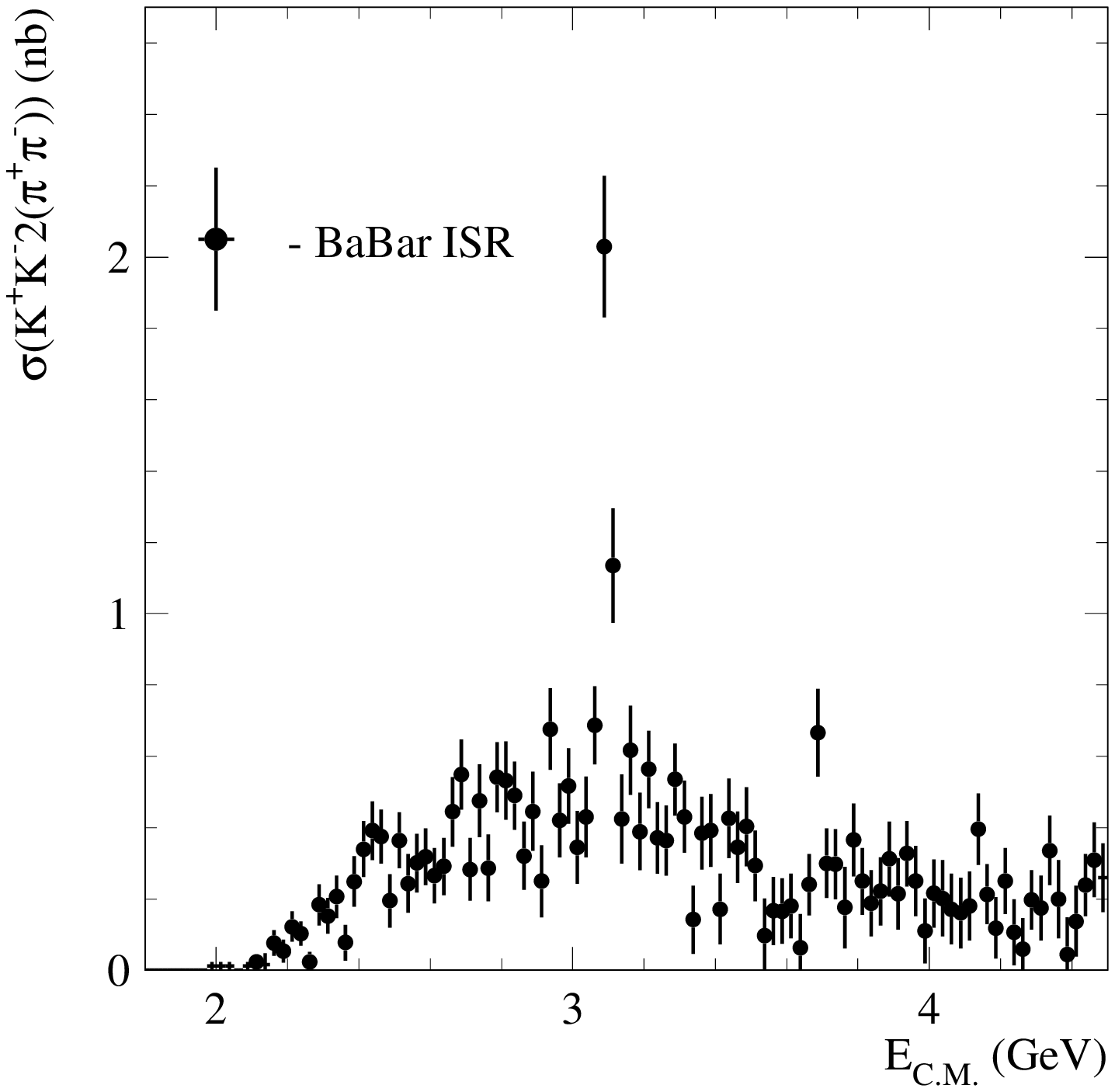} 
\caption{The $e^+e^-\to  K^+K^-\pi^+\pi^-\pi^+\pi^- $ cross section.} 
\label{sk2pi4} 
\end{minipage}
\end{figure}

{\bf SU(3) and QCD tests for baryon form factors.} 
New \babar\  data on baryon form factors give a possibility to compare 
them with the predictions from the 
known form factors models. A fit with the asymptotic QCD 
fitting function \cite{QCD} 
\begin{equation}
F\sim \alpha_S^2(m^2)/m^4\sim C/m^4ln^2(m^2/\Lambda^2),
\label{qcdf}
\end{equation}
applied to the proton form factor data, is shown in Fig.\ref{qcdprot}. Here
$\Lambda=0.3~GeV$, and C is a free parameter. If we neglect the steps at 2.15 and
2.9 GeV, the fit (\ref{qcdf}) in Fig.\ref{qcdprot} describes the data fairly 
well, indicating the asymptotic behaviour already at $2\div3$ GeV.

  The same fit (\ref{qcdf}) for  the $\Lambda$ form factor            
is not so good (see the curve labeled $n=4$ in the Fig.\ref{qcdlam}).
The fit improves  if we take  power $m^8$ in the                    
denominator of Eq.\ref{qcdf} instead of  $m^4$
(Fig.\ref{qcdlam}, curve $n=8$). 
The same results hold in the fitting of the 
$\Sigma^0$  and $\Sigma^0\Lambda$ formfactors. We conclude that  
$\Lambda$, $\Sigma^0$ and $\Sigma^0\Lambda$ form factors are considerably 
steeper  than the proton form factor.

  In SU(3) symmetry model the octet baryon form factors are related to each
  other. The asymptotic predictions \cite{VChern} are:      
  $F_{n}=1.94F_{\Lambda},~~ F_{p}=2.13F_{n},~~F_{\Sigma^0}=-1.18F_{\Lambda},              
  ~~F_{p}=4.1 F_{\Lambda}$.               
  A test of these predictions in  Fig.\ref{netrlamb} for  the doubled \babar\
$\Lambda$ \cite{LLBar} and Fenice neutron  \cite{Fenice}  form factors 
at 2.4 GeV shows good agreement with prediction $F_{n}=1.94F_{\Lambda}$.
This result is important for the planned 
neutron form factor measurement \cite{SND}.              
The comparison of the proton              
  and  $\Lambda$ form factors in Fig.\ref{protlamb}, shows that the data at    
  $E<3$ GeV are far from the asymptotic 
QCD prediction $F_{p}=4F_{\Lambda}$. But the course
of  both form factors with energy indicates  that 
  above 4 GeV the agreement with the QCD predictions is expected.

{\bf Final states including kaon  pairs.}
The ISR approach is also applied to study   $e^+e^-$
annihilation cross sections with a pair of kaons  in the final state. 
Figure \ref{kkspi} shows the $e^+e^-\to K_SK^{\pm}\pi^{\mp} $   
cross section \cite{k2pi} with a  peak at $\sim 1.7~GeV$ 
mainly from the $\phi^{\prime}(1680)$ state. 
The Dalitz plot in Fig.\ref{dalkks} shows  
that $KK^{\star}(892)$ and $KK_2^{\star}(1430)$ intermediate states 
dominate in the $K\bar{K}\pi$ final state.  
The fitting of the  $e^+e^-\to K\bar{K}\pi$
cross sections with  the sum of the expected contributions
from  the $\phi$,  $\phi^{\prime}$,  $\phi^{\prime\prime}$,
$\rho^0$,  $\rho^{\prime}$,  $\rho^{\prime\prime}$ state is done. 
The parameters of  the $\phi^{\prime}$ and other excited vector meson 
states, obtained from the fit,
are compatible with their PDG  values. 

 In the  cross section of the $e^+e^-\to\phi\eta,~ \eta\to\gamma\gamma$ 
 process (Fig.\ref{sigphieta}) the peak from the  $\phi^{\prime}(1680)$ 
is seen. Another small              
peak is observed  with $M=2139\pm 35$ MeV,            
$ \Gamma=76\pm 62$ MeV and $2\sigma$ significance.
In general  the $e^+e^-\to\phi\eta$ channel is 
very suitable for a search of
$\phi^{\prime}$s states,  because it is entirely          
isoscalar and a contribution of  $\omega^{\prime}$s states is            
OZI suppressed. Another channel $e^+e^-\to \phi\pi^0 $ 
is potentially  suitable  for a  search of exotic states,
because the ordinary vector mesons decays into $\phi\pi^0 $     
are strongly OZI suppressed. The $e^+e^-\to \phi\pi^0 $ cross section
is measured for the first time  \cite{k2pi} and 
found to be very small $\leq 0.1~nb$.

   The measured 
$e^+e^-\to K^+K^-\pi^+\pi^-$ and $e^+e^-\to K^+K^-\pi^0\pi^0$ 
cross sections \cite{k2pi2} are of the order of several nanobarn  
with  an enhancement at 1.7 GeV. In the final state mode $\phi f_0(980)$, 
$f_0(980)\to \pi^+\pi^-, \pi^0\pi^0$ a peak is observed with 
$M=2175\pm18$ MeV, $\Gamma=58\pm2$ MeV and $\Gamma_{ee}\simeq 2.5$ eV.
The new state is named $Y(2175)$. It is even more distinctly seen in the 
final state $K^+K^-f_0$  (Fig.\ref{x2175}).
The nature of $Y(2175)$ is not yet clear. It might be the 
$\phi^{\prime\prime}$ state, a four-quark or molecular $(ss\bar{s}\bar{s})$ 
state or a light analogue of the known $Y(4260)$, 
because they are both relatively narrow and have close electron widths
($\Gamma_{ee}(Y(4260))\simeq 5.5$ eV). 
If the peak at 2139 MeV  in the $\phi\eta$ channel  mentioned above  
is another decay channel of $Y(2175)$, then its electron width
should be larger than the quoted 2.5 eV value.


  Several new  channels with a pair of kaons are studied at Babar
\cite{k2pi2,k2pi3,k2pi4}: $e^+e^-\to K^+K^-K^+K^-,~
K^+K^-\pi^+\pi^-\pi^0,~K^+K^-\pi^+\pi^-\pi^+\pi^-,~
K^+K^-\pi^+\pi^-\eta$. In the $e^+e^-\to 4K$ process the $\phi K^+K^-$                   
intermediate state is dominant. 
In the $e^+e^-\to K^+K^- 3\pi$ cross section the 
$\omega (783)$  and $\eta (550)$ are clearly seen in the $3\pi$ mass spectrum.
As an example, the $e^+e^-\to K^+K^-4\pi$  cross section is shown in            
Fig.\ref{sk2pi4}. In general, the substructures in                                 
all final states with kaons deserve a more careful study.

{\bf Estimation of strangeness contribution to the total hadronic 
cross  section}. The  quantity $R=\sigma(e^+e^-\to hadrons)/\sigma(e^+e^-\to
\mu^+\mu^-)$, important in   low energy physics, 
below the charm threshold consists of the contributions of $u,d,s$
quarks. For the strange quark the relative contribution is 1/6.
To calculate $R$,  the direct cross section measurements
are preferable, because the QCD calculations in the non asymptotic region 
can be not  sufficiently precise.
Based on  the \babar\ data reported here
the total strangeness cross section 
is summarized at two points 2.5 and 3 GeV and found to be equal 
to 3.3 and 2.1 nb, respectively. This would be  compared with
the expected   5 and 3.5 nb of the strangeness  cross section 
and 28 and 19 nb of
the total hadronic cross section. So the measured 
cross sections constitute about 2/3 of the  strangeness cross section 
and 10\% of the total hadronic cross section.
The rest  channels,  such as  $K_S K_L\pi$, $K^{\pm}K_S\pi\pi$,   
$K_SK_S\pi\pi$ and others can be also measured 
using the initial state radiation. 

{\bf ISR perspectives}. In a few   years 
the total integrated luminosity at  B-factories  
is expected to reach  $\sim 2~ab^{-1}$, that  
is about 10 times larger than used in the present analysis. 
This gives  hope for more accurate measurements
of the reactions considered here
and extending the
measurements to larger masses.

{\bf Conclusions}. Using the initial state radiation at \babar\, 
the cross sections $e^+e^-\to \proton\antiproton,~ \Lambda\Lbar,~
\Sigma^0\Sigbar^0,~ \Lambda\Sigbar^0(\Sigma^0\Lbar)$ are measured. 
The measured baryon timelike form factors 
are compared with model 
predictions. The cross sections with a pair of kaons
$e^+e^-\to  K\bar{K}\pi (\eta),~K^+K^-\pi\pi, ~K^+K^-3\pi,$
$K^+K^-4\pi,$ $K^+K^-K^+K^-$
are also measured. In the $K^+K^- f_0(980)$  final state,  a  new
state $Y(2175)$ with $M=2175\pm18$ MeV,  $\Gamma=58\pm2$ MeV is observed. The
total measured strangeness cross section  
constitutes about 2/3 of the full strangeness cross section  
and $\sim 10\%$ of the total hadronic cross section.

{\bf Acknowledgment.}
 The author is grateful  for fruitful discussions to
Vladimir Druzhinin and Victor Chernyak.

\end{document}